\documentclass[prl,preprint,endfloats]{revtex4-1}
\usepackage{bm,graphicx,graphics,amsmath,amssymb,bm,epsfig,color}
\usepackage{euscript,tabularx}
\usepackage{longtable}

\begin{document}

\title{Detection of the microwave spin pumping using the inverse spin Hall
  effect}

\author{C. Hahn}
\author{G. de Loubens}
\author{M. Viret}
\author{O. Klein}\email[Corresponding author:]{ oklein@cea.fr}

\affiliation{Service de Physique de l'\'Etat Condens\'e (CNRS URA
  2464), CEA Saclay, 91191 Gif-sur-Yvette, France}

\author{V.V. Naletov}

\affiliation{Service de Physique de l'\'Etat Condens\'e (CNRS URA
  2464), CEA Saclay, 91191 Gif-sur-Yvette, France \\ Institute of
  Physics, Kazan Federal University, Kazan 420008, Russian Federation}

\author{J. Ben Youssef}

\affiliation{Universit\'e de Bretagne Occidentale, Laboratoire de
  Magn\'etisme de Bretagne CNRS, 6 Avenue Le Gorgeu, 29285 Brest,
  France}

\begin{abstract}
  We report on the electrical detection of the dynamical part of the
  spin pumping current emitted during ferromagnetic resonance (FMR)
  using the inverse Spin Hall Effect (ISHE). The experiment is
  performed on a YIG$|$Pt bilayer. The choice of YIG, a magnetic
  insulator, ensures that no charge current flows between the two
  layers and only the pure spin current produced by the magnetization
  dynamics is transferred into the adjacent strong spin-orbit Pt layer
  via spin pumping. To avoid measuring the parasitic eddy currents
  induced at the frequency of the microwave source, a resonance at
  half the frequency is induced using parametric excitation in the
  parallel geometry. Triggering this nonlinear effect allows to
  directly detect on a spectrum analyzer the microwave component of
  the ISHE voltage. Signals as large as 30~$\mu$V are measured for
  precession angles of a couple of degrees. This direct detection
  provides a novel efficient means to study magnetization dynamics on
  a very wide frequency range with great sensitivity.
\end{abstract}

\maketitle


One great expectation of spintronics regarding information technology
is the promise that pure spin currents can be generated and
manipulated without their charge current counterparts
\cite{jungwirth12}. Pure spin currents correspond to the transport of
angular momentum in a very wide range of materials including metals
and insulators with or without magnetic order. In ferromagnetic
metals, charge currents are intrinsically associated to spin currents
because electrons at the Fermi level are spin polarized. Using these
as injection electrodes, pure spin currents can be generated into a
non magnetic metal in a non-local geometry where charges are evacuated
through one electrode whereas spin diffusion can be collected by
another nearby electrode \cite{valenzuela06,kimura07}. This lateral
geometry is well suited to nanostructures, but it is limited by the
required current densities and the short spin diffusion lengths
\cite{otani08}. Another option relies on using the spin Hall effect, a
phenomenon based on the spin-orbit interaction of a charge current
which generates a transverse spin current in a conductor
\cite{dyakonov71,hirsch99}. Pure spin currents can also be generated
in ferromagnetic insulators by the spin pumping mechanism
\cite{tserkovnyak05,woltersdorf05,heinrich11} during magnetization
precession. This effect is produced by the damping of spin-waves which
transfer angular momentum across an interface to a neighbouring layer.
The emitted pure spin current can be detected electronically in an
adjacent layer by the inverse spin Hall effect (ISHE) using metals
with strong spin-orbit coupling like Pt
\cite{saitoh06,ando08,ando09,mosendz10,mosendz10a}. The novelty here
offered by electrical detection of the spin pumping using the ISHE is
that it can be used also on non-metallic ferromagnets, including
Yttrium Iron Garnet (YIG)
\cite{kajiwara10,heinrich11,wang11,sandweg11,kurebayashi11,vilela-leao11,chumak12,castel12a,hahn13},
a magnetic insulator which has unsurpassed small damping in
ferromagnetic resonance (FMR). But so far, only the dc component of
the ISHE voltage induced by FMR has been measured, which is a second
order effect in the precession angle. Here, we report on a direct
measurement of its first order ac counterpart.

\begin{figure}
  \includegraphics[width=15cm]{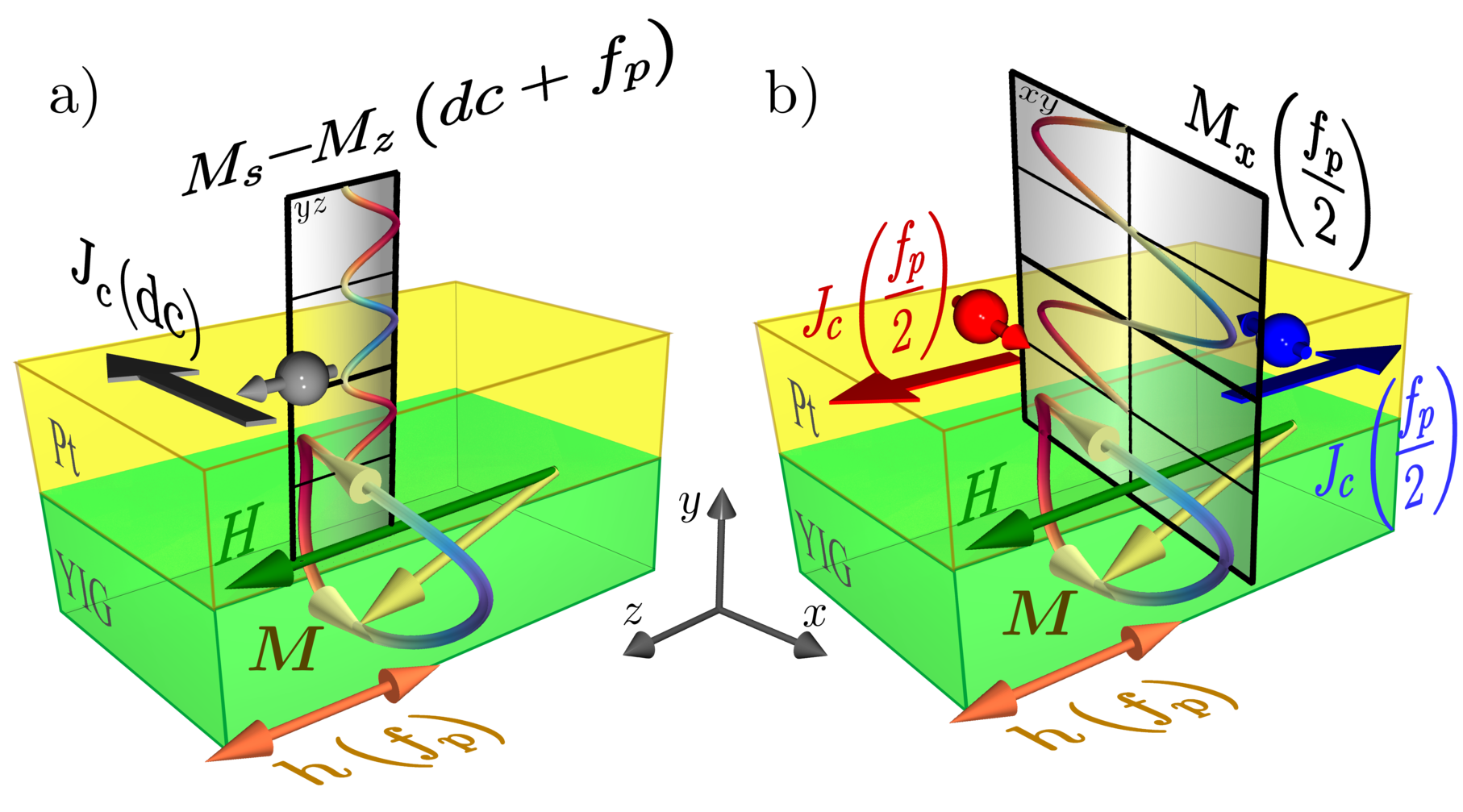}
  \caption{(Color online) Schematic representation showing the
    direction of the dc (a) and ac (b) charge currents produced when a
    pure spin current is pumped from the insulating magnetic YIG
    (green) into the strong spin-orbit Pt metal (yellow). The
    instantaneous magnetization ${\bm M}(t)$ is shown in a bi-variate
    colormap: the blue-red colors code the $x$-component and the gray
    shades code the $z$-component.  The precession of ${\bm M}$ at
    $f_{p}/2$ around ${\bm z}$, is driven by parametric excitation: it
    requires the pumping field $h$, oscillating at $f_p$, to be
    parallel to the bias magnetic field $H$, and the precession of
    $(M_x, M_y)$ to be elliptic
    (thus $M_x^2+M_y^2$ is not a constant of the motion). The
    flowcharts on top illustrate that $M_z=\sqrt{M_s^2 -M_x^2-M_y^2}$
    then oscillates twice faster than $M_x$ or $M_y$. Spin pumping
    currents are flowing from YIG to Pt (\textit{i.e.} along the
    $y$-axis). The injected angular momentum in Pt is carried by the
    spins (arrows attached to the electrons opposite to their angular
    momentum).  The direction of the instantaneous charge current
    (flat arrow) is given by the right hand rule, see
    Eq.(\ref{eq:jejs}).}
  \label{FIG1}
\end{figure}

The experiments of the present study are performed at room temperature
on a YIG$|$Pt bi-layer where the YIG is a 200~nm thick epitaxial film
grown by liquid phase epitaxy. A 6~nm thick Pt layer is then sputtered
on top and two contact electrodes are defined at each end. The sample
is mounted on a stripline antenna generating a microwave field $h$
oscillating at a frequency $f_p$ as sketched in Fig.\ref{FIG1}. At
resonance of the uniform mode, the YIG emits, perpendicularly to the
YIG$|$Pt interface, a flow of angular momentum generated by the spin
pumping effect,
\begin{equation} 
  {\bm J}_s = \left ( \frac{\hbar}{2eM_s} \right )^2 G_{\uparrow \downarrow}
  \left [\bm M \times \frac{\partial \bm M}{\partial t} \right] \ .
\label{eq:Js}
\end{equation}
In this expression, $\bm M$ is the magnetization vector, whose norm is
$M_s$, $\hbar$ is the reduced Planck constant, $e$ the electron
charge, and $ G_{\uparrow \downarrow}$ the spin mixing conductance at
the YIG$|$Pt interface in units of $\Omega^{-1}\cdot$m$^{-2}$. The
spin current pumped into the adjacent Pt is then converted into a
charge current by ISHE,
\begin{equation}
  {\bm J}_c=\frac{2e}{\hbar}\Theta_\text{SH}\left[{\bm y} \times {\bm J}_s \right] \, ,
  \label{eq:jejs}
\end{equation}
where $\Theta_\text{SH}$ is the spin Hall angle in Pt and $\bm y$ is
the unit vector perpendicular to the interface (see Fig.\ref{FIG1}).

Importantly, the flow $ {\bm J}_s$ (and hence $ {\bm J}_c$) has both
dc and ac components \cite{jiao13}. The dc part of this signal is
normally detected as a voltage, proportional to ${\bm
  J}_{c}(\text{dc})$, that is maximum in a transverse geometry
(\textit{i.e.} $\perp H$, see Fig.\ref{FIG1}a). In contrast, ${\bm
  J}_{c}(\text{ac})$ is maximum in a parallel geometry (\textit{i.e.}
$\parallel H$, see Fig.\ref{FIG1}b). It is interesting to note that
for circular precession the dc signal is second order in the
precession angle $\theta$ ($\propto \sin^2\theta$, maximal for
$\theta=90^\circ$), while its ac counterpart is first order ($\propto
\sin \theta \cos \theta$ and maximal for $\theta=45^\circ$). Thus the
ratio of ac to dc scales as $1/ \tan \theta$, which is large for small
precession angles. However, the ac component is much harder to detect
as it oscillates obviously at the same frequency as that of the
microwave generator producing the FMR. This microwave excitation field
$h$ induces eddy currents in any closed circuit containing the
sample. These spurious ac currents are generally rather large and
dominate any other contribution at the same frequency. Therefore, a
clear detection of the ac spin currents has not yet been successful as
one has to carefully eliminate the large amplitude eddy currents. In
this letter, we report on the unambiguous detection of these ac spin
currents emitted at ferromagnetic resonance using a specially designed
system leading to an ac signal totally unpolluted by any other
contribution. The key strategy here is to generate the FMR at half the
frequency of the excitation source. This phenomenon is known as
parametric excitation \cite{sparks64}. It exploits the fact that due
to the ellipticity of the in-plane precession, the magnetization
follows a clamshell trajectory. During a full revolution of ${\bm M}$
around its precession axis $z$, the $z$-component of the magnetization
$M_z$ oscillates twice faster, see Fig.\ref{FIG1}a. This is also
illustrated in Fig.\ref{FIG1}b using red and blue colors to code its
$x$-component, $M_x$. Therefore, by exciting parallel to ${\bm M}$,
one can trigger the precession at half the source frequency. One
should note however that this parallel parametric excitation is only
possible in systems with low damping, since the excitation power needs
to exceed a minimum threshold corresponding to a fraction of linewidth
(typically below one Oersted for YIG \cite{sparks64}) in order to
drive the magnetization into oscillation.

Experimentally, since the technique depends sensitively on the
respective orientations of the microwave excitation and the bias
field, we shall measure the ratio of the dc and ac components of ${\bm
  J}_c$, by rotating $H$, the static bias magnetic field, in the film
plane. For all practical purposes, the YIG slab can be considered as
an infinite film, whose resonance conditions are independent of the
orientation of $H$.

\begin{figure}
  \includegraphics[width=15cm]{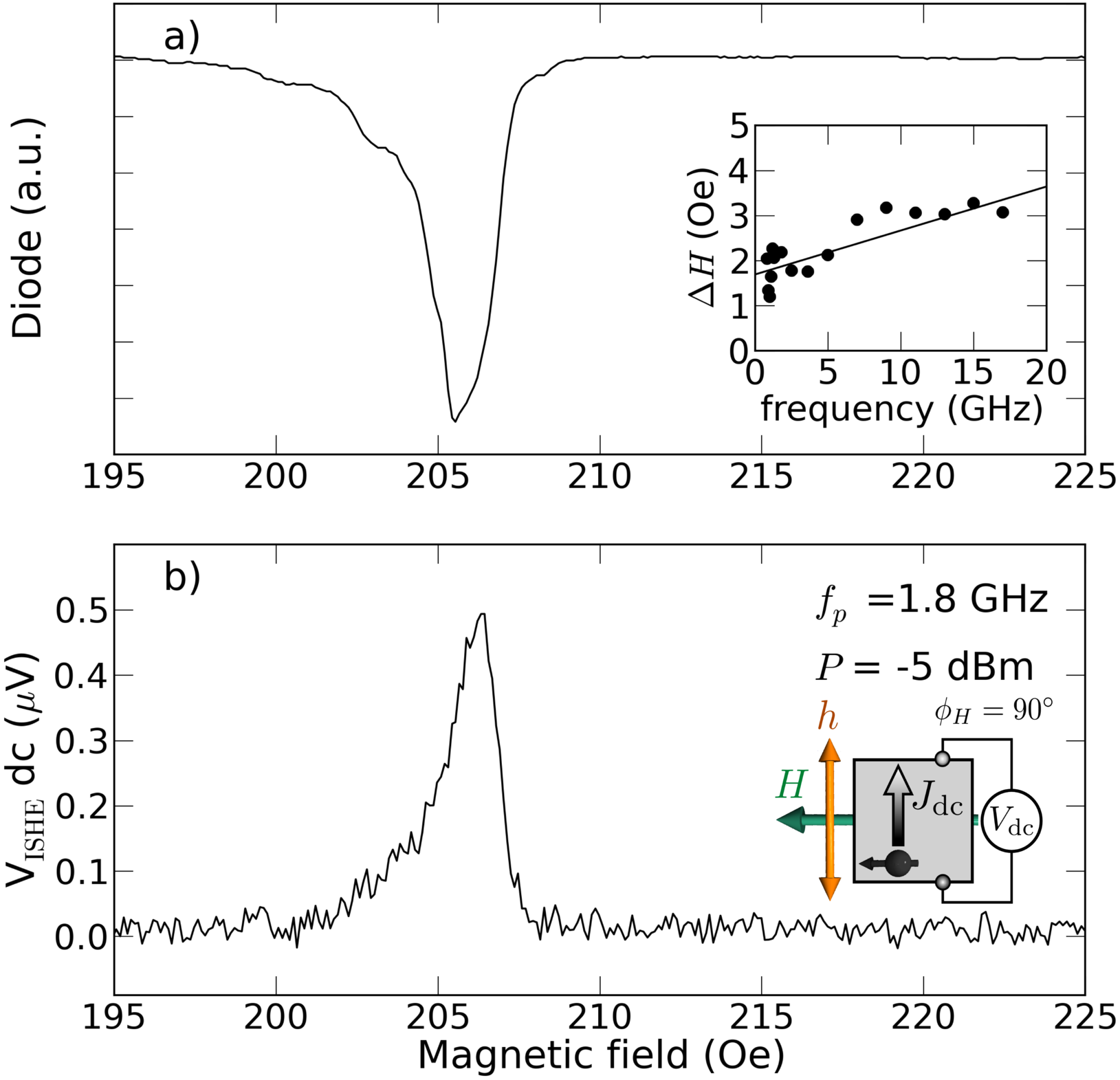}
  \caption{(Color online) Standard FMR detected using dc ISHE voltage. The pumping
    field $h$ oscillating at $f_p=1.8$~GHz is oriented perpendicularly
    to the static magnetic field $H$. (a) Larmor absorption peak of
    the uniform mode detected with a diode (the inset shows the
    dependence of the linewidth on frequency). (b) Corresponding dc
    ISHE voltage measured perpendicularly to $H$ (the inset shows the
    geometry of the experiment).}
  \label{FIG2}
\end{figure}

\begin{figure}
  \includegraphics[width=15cm]{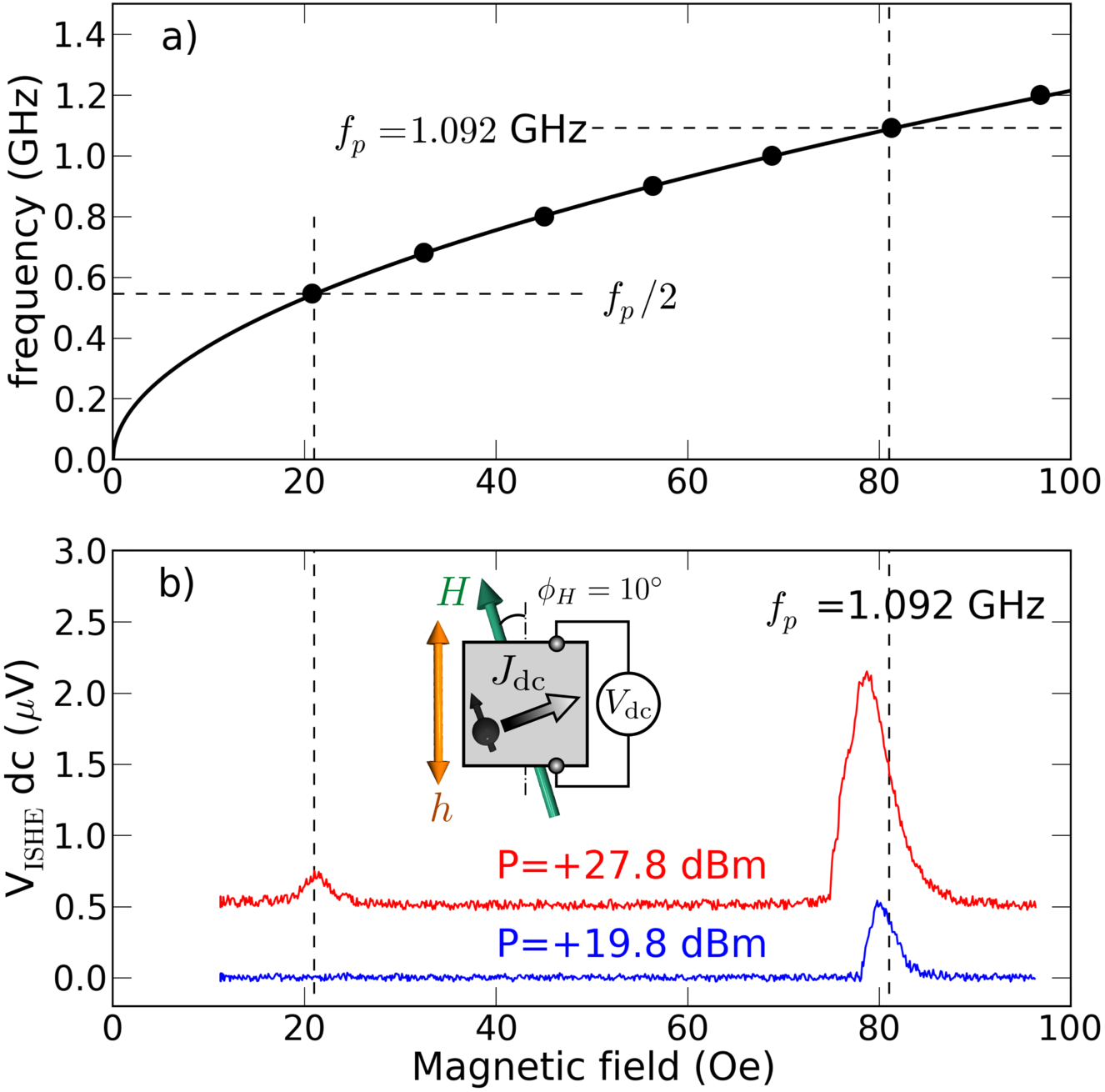}
  \caption{(Color online) Parametric excitation detected using the dc ISHE
    voltage. (a) Measured variation of the resonance frequency
    $f_\text{res}$ as a function of the applied field (dots) using the
    standard FMR geometry (Fig.\ref{FIG2}). The solid line is the
    Kittel law $f_\text{res} = (\gamma/2\pi)\sqrt{H(H+4\pi M_s)}$ with
    $\gamma=1.785\cdot10^7$~rad$\cdot$s$^{-1}\cdot$Oe$^{-1}$ and
    $M_s=139$~emu$\cdot$cm$^{-3}$. (b) Spin wave mode spectra detected
    using the dc ISHE voltage for two different power levels. The bias
    field $H$ is oriented at 10$^\circ$ from the pumping field $h$
    oscillating at $f_p=1.092$~GHz (the inset shows the geometry of
    the experiment). At lower power (+19.8 dBm), the only peak
    detected in the spectrum occurs when the Larmor condition is met
    at $f_p$. At higher power (+27.8 dBm), a new peak appears in the
    spectrum at $f_p/2$, corresponding to the parametrically excited
    resonance.}
  \label{FIG3}
\end{figure}

In order to characterize our sample, and in particular the electrical
conversion of the pumped spin current at the YIG$|$Pt interface, we
first perform standard FMR resonance, where the small microwave field
$h$ is perpendicular to $H$ ($\phi_H=90^\circ$). It is indeed the most
efficient configuration to excite the magnetization dynamics: in the
case displayed in Fig.\ref{FIG2}, the angle of precession induced in
YIG at resonance by a microwave field $h=36$~mOe ($P=-5$~dBm) is
$\theta=1.1^\circ$ (see Supplementary Materials). The FMR signal is
detected simultaneously by probing the power transmitted through the
microwave line using a diode (Fig.\ref{FIG2}a) and by measuring the dc
ISHE voltage transversally to the static magnetization
(Fig.\ref{FIG2}b). Both measurements yield the same evolution of the
resonance field versus frequency following the Kittel law for an
in-plane magnetized thin film, see Fig.\ref{FIG3}a. By measuring the
diode signal at low power ($P=-20$~dBm, corresponding to $h=6$~mOe),
one can also determine in the linear regime the dependence of the
linewidth on frequency, which is reported in the inset of
Fig.\ref{FIG2}a. A linear fit yields the Gilbert damping,
$\alpha_G=(1.4\pm0.1)\cdot 10^{-4}$, highlighting the very small
magnetic relaxation of our YIG film \cite{hahn13}. The inhomogeneous
part to the linewidth, $\Delta H_0=1.7\pm0.2$~Oe, reflects sample
imperfections specific to the growth process of this batch. We find
that this contribution dominates the broadening below 10~GHz. The
amplitude of $V_\text{ISHE}(\text{dc})$ measured at resonance allows
us to determine the transport parameters at play in the electrical
conversion of the pumped spin current (see Supplementary
Materials). We find that our dc ISHE data can be well explained using
typical parameters of the YIG$|$Pt system \cite{castel12a,hahn13}:
spin diffusion length $\lambda_\text{sd}=2$~nm, spin Hall angle
$\Theta_\text{SH}=0.05$, and spin mixing conductance $G_{\uparrow
  \downarrow} = 10^{14}~\Omega^{-1}\cdot$m$^{-2}$.

We now wish to excite parametrically the YIG magnetization at half the
applied microwave frequency, $f_p/2$, by taking advantage of the
clamshell shape of the precession trajectory, which is sketched in
Fig.\ref{FIG1}. For this, the component of the microwave field $h$
parallel to the bias field $H$ should reach the excitation threshold
for parallel parametric excitation \cite{sparks64}. To demonstrate
this effect in our sample through dc ISHE voltage measurements, we set
a finite angle $\phi_H=10^\circ$ between $h$ and $H$ (see inset of
Fig.\ref{FIG3}b). Compared to the previous perpendicular geometry, the
resonance condition at $f_p$ has not changed, only the microwave field
is less efficient to bring the magnetization out-of-equilibrium, thus
a stronger excitation power should be used to reach the same
precession angle. We move momentarily to lower frequency in order to
insert in the microwave circuit an additional amplifier limited in
bandwidth to 1.1~GHz (see Supplementary Materials). At $f_p=1.092$ GHz and
$P=+19.8$~dBm we observe only one resonance peak at $H=80$~Oe in
Fig.\ref{FIG3}b. The new feature here is that if we increase the power
to $P=+27.8$~dBm, a second peak appears in the spectrum at $H=20$
Oe. Looking at the Kittel law of Fig.\ref{FIG3}a), we find that it
corresponds to the uniform mode resonating at $f_p/2$, which is thus
parametrically excited
\cite{sandweg11,kurebayashi11,kurebayashi11a,ando12}. Thanks to the
quantitative analysis of $V_\text{ISHE}(\text{dc})$ using the
transport parameters determined previously, it is possible to quantify
the angle of precession corresponding to this parametric excitation:
$\theta=5.1^\circ$ (see Supplementary Materials).

\begin{figure}
  \includegraphics[width=15cm]{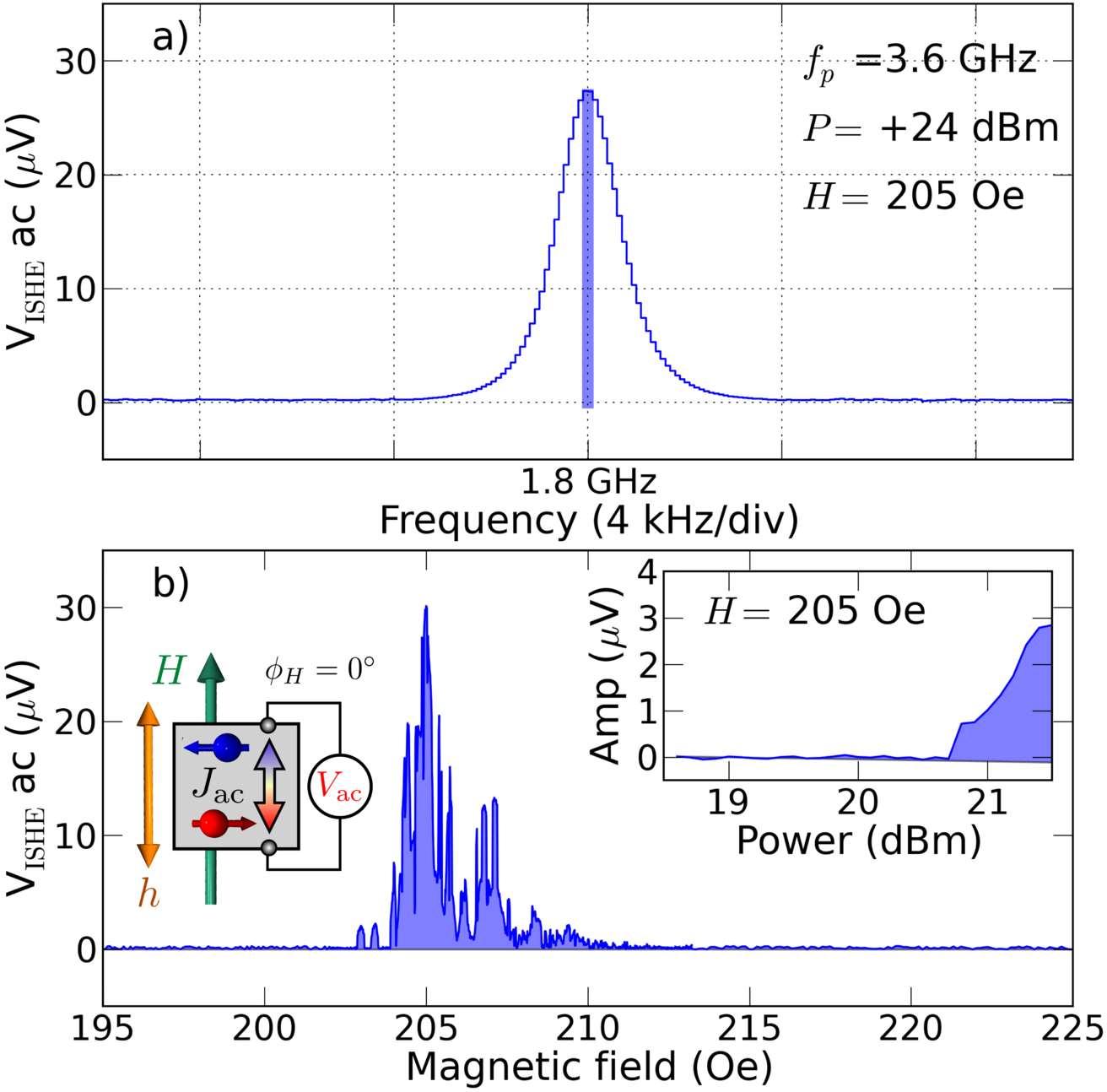}
  \caption{(Color online) Parametric excitation detected using the ac
    ISHE voltage. The large power (+24~dBm) pumping field $h$ at
    $f_p=3.6$~GHz is oriented parallel to the static magnetic field
    $H$. (a) The ac ISHE voltage generated by the parametric
    excitation at $H=205$~Oe is monitored on a spectrum analyzer
    (1~kHz resolution bandwidth): an oscillation voltage is detected
    at $f_p/2=1.8$~GHz. (b) The amplitude of this oscillation is
    measured as the bias field is swept from 195 to 225~Oe. The
    envelope of the curve should be compared to that in
    Fig.\ref{FIG2}. The maximum parametric signal occurs at $H_{\rm
      res}=205$~Oe. The inset shows the threshold behavior ($P_c
    \simeq +20.7$~dBm) of the power dependence of the spectrum
    analyzer signal at 1.8~GHz to demonstrate the parametric
    excitation.}
  \label{FIG4}
\end{figure}

\begin{figure}
  \includegraphics[width=15cm]{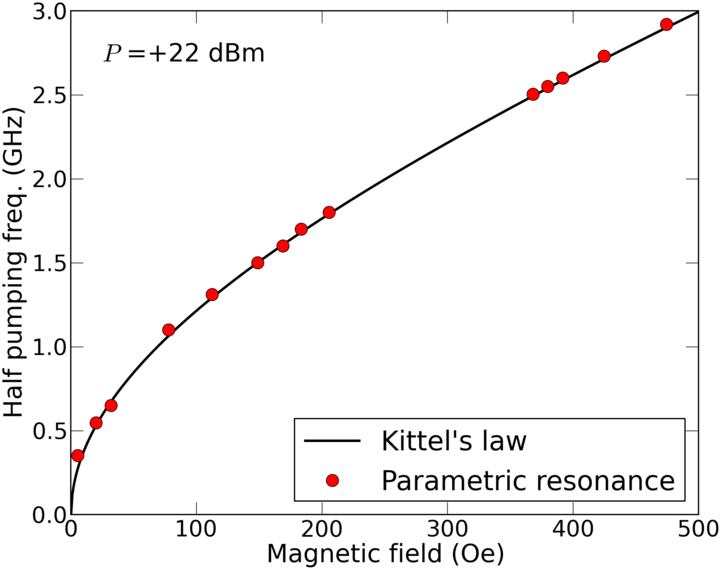}
  \caption{(Color online)  voltage at half the pumping frequency. Red dots correspond
    to the bias field required to observe the maximum parametric
    signal detected on the spectrum analyzer at $f_p/2$. The solid
    line is the Kittel law of the YIG$|$Pt bilayer (see
    Fig.\ref{FIG3}a).}
  \label{FIG5}
\end{figure}

The next step is to directly detect the ac ISHE voltage generated at
$f_p/2$ by the parametrically excited magnetization dynamics. For
this, we now align the microwave field $h$ with the bias field $H$
($\phi_H=0^\circ$) and excite the system at $f_p=3.6$~GHz and high
power, $P=+24$~dBm. The two voltage leads which contact the Pt layer
are connected directly into a spectrum analyzer (SA) without any
preamplification scheme. By sweeping the frequency of the SA at fixed
$H=205$~Oe, we detect a large signal of amplitude $30~\mu$V at exactly
$f_p/2=1.8$~GHz, as can be seen in Fig.\ref{FIG4}a. We claim this
signal to be the ac component of the pure spin current pumped from the
YIG parametric excitation into Pt, and converted into a voltage by
ISHE. To prove this, we plot in Fig.\ref{FIG4}b the amplitude of the
SA signal measured at $f_p/2$ as a function of $H$. We find that the
amplitude of the signal is maximum at $H=205$~G, which is the
resonance field determined by standard FMR at 1.8~GHz in
Fig.\ref{FIG2}, and dies out in a range of about $\pm1.5$~Oe around
this field. We have also checked that this ac voltage signal has a
parametric excitation origin, by studying its amplitude as a function
of the excitation power. One can observe in the inset of
Fig.\ref{FIG4}b that the peak at $f_p/2$ suddenly appears on the SA
above a critical power $P_c=+20.7$~dBm, \textit{i.e.}, a critical
microwave field $h_c=0.7$~Oe, in good agreement with the expected
threshold for parallel parametric excitation in YIG
\cite{sparks64}. We note that the envelope of the parametric
excitation signal observed in Fig.\ref{FIG4}b as a function of $H$ has
a shape close to that of the standard FMR peak of
Fig.\ref{FIG2}. Still, we observe abrupt jumps of the amplitude as $H$
is varied. We find that the details of this variation are very
sensitive to changes in the orientation $\phi_H$ of the bias field. We
attribute this to the excitation of spin-wave modes which are almost
degenerate with the uniform precession mode
\cite{sparks64,naletov07}. We have repeated the same experiment for
different excitation frequencies ranging from 0.8~GHz up to
6~GHz. Because the microwave power exceeds the threshold for parallel
parametric excitation ($P=+22$~dBm), an ac ISHE voltage at half the
excitation frequency is observed on the SA as a function of $H$ when
the condition for resonance is met at $f_p/2$ (cf. Kittel law on
Fig.\ref{FIG5}). Lastly, in order to check experimentally that our
method eliminates \emph{all} spurious signals, we performed the same
series of measurements on a reference sample where the Pt layer was
replaced by 15~nm of Al. We found experimentally that \emph{no}
electrical signal is produced at $f_p/2$, which allows to conclude
that the layer with strong-spin orbit scattering like Pt is
indispensable to observe the ac ISHE voltage.

To gain more insight into the ac ISHE voltage, we would like to
comment on its amplitude. For that, we compare the ac and dc
components using the $\phi_H=10^\circ$ configuration (see
Fig.\ref{FIG3}b). We measure an effective dc voltage
$V_\text{ISHE}(\text{dc})=1.1~\mu$V, while in the same conditions, the
generated ac voltage is $V_\text{ISHE}(\text{ac})=6.2~\mu$V (see
Supplementary Materials). Thus, we obtain the experimental ratio $
\left[ V_\text{ISHE}(\text{ac})/V_\text{ISHE}(\text{dc}) \right
]_\text{exp}=5.6$. We have determined that the angle of precession
corresponding to the parametric excitation observed at $H=20$~Oe is
$\theta=5.1^\circ$. Therefore, one would expect that $ \left
  [J_{s}(\text{ac})/J_{s}(\text{dc}) \right ]_\text{theo} =
\xi/\tan{\theta}=5.7$, where $\xi=0.51$ is the ellipticity correction
factor for this case (see Supplementary Materials), in excellent
agreement with our estimation above. 

In conclusion, we have shown that the microwave part of the spin
pumping current emitted by a ferromagnet driven at resonance can be
detected by the inverse spin Hall effect using an adjacent metallic
layer with strong spin-orbit scattering \footnote{During this work, we
  became aware of a parallel effort reporting also detection of ac-ISHE
  but in NiFe$|$Pt: http://arxiv.org/abs/1307.2961}. On our
millimeter size YIG$|$Pt sample it leads to a micro-Volt range
microwave signal measurable directly on a spectrum analyzer without
any pre-amplification or impedance matching. We believe that this
broad band direct detection provides a novel efficient means to study
magnetization dynamics in a wide variety of ferromagnetic
materials. Further analysis of the measured spectra in the parallel
parametric geometry will provide new insights into the spin-wave
competition in the nonlinear regime. The phenomenon will also allow
dynamical studies of the process of spin transfer at the interface
with strong spin orbit non-magnetic metals \cite{jia11,liu12,miron11}.

This research was supported by the French ANR Grant Trinidad (ASTRID
2012 program) and by the RTRA-2011 grant Spinoscopy. 

\appendix

\section{Supplementary Materials: Sample preparation and characterization}

The 200~nm thick single crystal Y$_3$Fe$_5$O$_{12}$ (YIG) film was
grown by liquid phase epitaxy on a (111) Gd$_3$Ga$_5$O$_{12}$ (GGG)
substrate. A 6~nm thin Pt layer with conductivity
$\sigma=(2.3\pm0.2)\cdot10^6~\Omega^{-1}\cdot$m$^{-1}$ was then
sputtered on top. Vibrating sample magnetometry reveals that the YIG
film does not exhibit any in-plane anisotropy and has both very low
coercivity (0.5~Oe) and saturation field (2~Oe). Its saturation
magnetization, $M_s=139$~emu$\cdot$cm$^{-3}$, is the one of bulk
YIG. This value was confirmed by in-plane broadband FMR (from 0.4~GHz
up to 17~GHz). FMR also allows to extract the gyromagnetic ratio,
$\gamma=1.785\cdot10^7$~rad$\cdot$s$^{-1}\cdot$Oe$^{-1}$.

\section{Experimental setup}

Experiments are performed at room temperature. The microwave cell is
placed at the center of an electromagnet which can be rotated. The
500~$\mu$m wide, 2~$\mu$m thick Au transmission line cell is connected
to a synthesizer providing frequencies up to 20~GHz and power up to
$+24$~dBm (additional boost to $+35$~dBm is obtained by using an
external amplifier for frequencies lower than 1.1~GHz).  A standard
diode provides a quadratic detection of the transmitted power. The
calibration of the linearly polarized in-plane microwave field yields
$h=60$~mOe when the output power is 0~dBm. The YIG$|$Pt sample is cut
into a slab with lateral dimensions of 1.7~mm $\times$ 7~mm and placed
on top of the transmission line. The sample is mounted upside-down on
the stripline with the Pt layer facing the Au antenna.  The two
voltage leads are equidistant from the area of excitation and
separated by 3~mm (the two-point resistance is
$R=129~\Omega$). $V_\text{ISHE}(\text{dc})$ is measured by a lock-in
technique with the microwave power turned on and off at a frequency of
1.9~kHz. $V_\text{ISHE}(\text{ac})$ is directly measured using a
spectrum analyzer (input impedance $Z_0=50~\Omega$). Due to impedance
mismatch, the voltage transmission coefficient from the YIG$|$Pt
sample to the spectrum analyzer is $T=1-\frac{R-Z_0}{R+Z_0}=0.56$. We
have checked that this value holds below 2~GHz using a network
analyzer.

\section{Analysis of the dc ISHE voltage}

The amplitude of the ISHE dc voltage generated in Pt by the
magnetization precession with an angle $\theta$ at frequency $f$ in
YIG is given by \cite{castel12a,hahn13}:
\begin{eqnarray}
  V_\text{ISHE}(\text{dc}) & = & \Theta_\text{SH} \frac{G_{\uparrow \downarrow}}{G_{\uparrow \downarrow}+\frac{\sigma}{\lambda_\text{sd}}\frac{1-\exp{(-2t/\lambda_\text{sd}})}{1+\exp{(-2t/\lambda_\text{sd})}}} \sin(\phi_H) \frac{\pi \hbar L \mathcal{E} f \sin^2(\theta)}{e t} \frac{(1-\exp{(-t/\lambda_\text{sd}}))^2}{1+\exp{(-2t/\lambda_\text{sd})}} \ .
  \label{eq:Vish}
\end{eqnarray}
In this expression, $G_{\uparrow \downarrow}$ is the spin mixing
conductance, $\sigma$ the conductivity of the Pt layer,
$\Theta_\text{SH}$ its spin Hall angle, $\lambda_\text{sd}$ its spin
diffusion length, $t$ its thickness, and $L$ the length of the sample
excited by the excitation field. $\mathcal{E}$ is a factor close to
unity, which depends on the ellipticity of the precession trajectory,
hence on the frequency $\omega/2\pi$ \cite{ando09}: in our case,
$\mathcal{E}=0.43$ at 0.546~GHz and $\mathcal{E}=1.06$ at 1.8~GHz.

The dc ISHE voltage is maximal when the angle $\phi_H$ between the
direction along which the voltage drop is measured and the static
magnetization equals 90$^\circ$, which is the case in our standard FMR
geometry (Fig.2). In this case, we have checked that the dc
ISHE voltage is odd in applied magnetic field, which shows that the
voltage generated at resonance is not due to a thermoelectrical
effect. The measured $V_\text{ISHE}(\text{dc})$ are in good agreement
with Eq.(\ref{eq:Vish}), using values estimated from high frequency
susceptibility for the precession angle $\theta$ and the following
typical transport parameters \cite{hahn13}: $\Theta_\text{SH} = 0.05$;
$\lambda_\text{sd}=2$~nm; $G_{\uparrow \downarrow} =
10^{14}~\Omega^{-1}\cdot$m$^{-2}$. For instance, at $f_p=1.8$~GHz and
$P=-5$~dBm, the angle of precession at resonance is estimated to be
$\theta=1.1^\circ$, which yields
$V_\text{ISHE}(\text{dc})=0.44$~$\mu$V, to be compared to the value of
0.5~$\mu$V measured in Fig.2b. The overall agreement
between Eq.(\ref{eq:Vish}) and the experimental measurements of
$V_\text{ISHE}(\text{dc})$ vs. power (from -20~dBm to +10~dBm) and
frequency (from 0.4 to 17~GHz) is within 30\%.

When the angle $\phi_H=10^\circ$, as in the parametric excitation
geometry of Fig.3b, the loss of sensitivity for the dc ISHE voltage
drops as $1/\sin{\phi_H}\simeq 5.7$. Experimentally, we measure that
at $P=+27.8$~dBm, the parametric excitation at $f_p/2=0.546$~GHz
generates a dc voltage $V_\text{ISHE}(\text{dc})=0.19~\mu$V (setting
$\phi_H=-10^\circ$ leads to a measured dc ISHE voltage of same
amplitude but opposite sign). Using the same parameters as for the
analysis in the standard FMR configuration, we can estimate the
corresponding angle of precession from Eq.(\ref{eq:Vish}):
$\theta=5.1^\circ$.

\section{Analysis of the ac ISHE voltage}

In order to calculate the ac spin current generated by an elliptical
precession trajectory, we follow the analysis performed to derive the
factor $\mathcal{E}$ in Eq.(\ref{eq:Vish}) for the dc ISHE voltage
\cite{ando09}. To do so, we use the Landau-Lifschitz-Gilbert equation
to compute at resonance the three components of the magnetization
precessing around its equilibrium axis $z$ and the expression of the
pumped spin current flowing accross the FM$|$NM interface (see Eq.(1)
of main text).

In the case of a circular precession, this yields
$J_{s,x}^\text{circ}(t)=-\omega \left ( \frac{\hbar}{2e} \right )^2
G_{\uparrow \downarrow} \sin{\theta} \cos{\theta} \sin{\omega t}$ and
$J_{s,z}^\text{circ}=\omega \left ( \frac{\hbar}{2e} \right )^2
G_{\uparrow \downarrow} \sin^2{\theta}$.  (In our geometry,
$J_{s,y}^\text{circ}(t)$ does not produce any charge current through
ISHE because its spin polarization is parallel to the normal $y$ of
the FM$|$NM interface, see Eq.(2) and Fig.1 of main text). Hence,
$|J_{s}^\text{circ}(\text{ac})|/J_{s}^\text{circ}(\text{dc})=1/\tan{\theta}$,
where $\theta$ is the angle of precession.

In the case of an elliptical precession, the same approach yields the
correction factors with respect to the circular case:
\begin{itemize}
\item
  $J_{s}(\text{dc})/J_{s}^\text{circ}(\text{dc})=\frac{2\omega(\omega_M+\sqrt{\omega_M^2+4\omega^2})}{\omega_M^2+4\omega^2}$,
  for the dc component of the pumped spin current. This is nothing
  else than the factor $\mathcal{E}$ appearing in Eq.(\ref{eq:Vish})
  \cite{ando09}.
\item
  $J_{s}(\text{ac})/J_{s}^\text{circ}(\text{ac})=\frac{2\omega}{\sqrt{\omega_M^2+4\omega^2}}$
  for its ac component.
\end{itemize}
In these expressions, $\omega_M=\gamma 4\pi M_s$. Hence, the ratio of
the ac and dc components writes:
\begin{equation}
  \frac{J_{s}(\text{ac})}{J_{s}(\text{dc})}=\frac{\sqrt{\omega_M^2+4\omega^2}}{\omega_M+\sqrt{\omega_M^2+4\omega^2}}\frac{1}{\tan{\theta}}=\frac{\xi}{\tan{\theta}} \, .
  \label{eq:acdc}
\end{equation}

Using the same transport parameters as for the analysis of the dc ISHE
voltage, one finds that the precession angle corresponding to the
$30~\mu$V ac ISHE voltage measured in Fig.4 at $f_p/2=1.8$~GHz equals
$\theta=2.5^\circ$. In this experiment, $\phi_H=0^\circ$, so that the
corresponding dc voltage vanishes (see Eq.(\ref{eq:Vish})).

Setting $\phi_H=10^\circ$, as in the experiment of Fig.3b, enables to
detect both the dc and ac voltages. As explained above, the angle of
precession estimated from the dc ISHE voltage at $f_p/2=0.546$~GHz is
$\theta=5.1^\circ$. It corresponds to an effective dc voltage
$V_\text{ISHE}(\text{dc})=1/\sin{\phi_H}\times0.19~\mu$V$=1.1~\mu$V. The
ac ISHE voltage measured in the same conditions is found to be
$3.5~\mu$V. This value has to be corrected due to the previously
mentioned impedance mismatch,
$V_\text{ISHE}(\text{ac})=1/T\times3.5~\mu$V$=6.2~\mu$V. Hence, the
experimental ratio between the ac and dc components of the ISHE
voltage current is
$[V_\text{ISHE}(\text{ac})/V_\text{ISHE}(\text{dc})]_\text{exp}=5.6$,
which is in very good agreement with the theoretical value from
Eq.(\ref{eq:acdc}),
$[J_{s}(\text{ac})/J_{s}(\text{dc})]_\text{theo}=5.7$.


%

\end{document}